# On the Applicability of Explainable Artificial Intelligence for Software Requirement Analysis

Behnaz Jamasb, Reza Akbari, and Seyed Raouf Khayami

{b.jamasb, akbari, khayami}@sutech.ac.ir

Department of Computer Engineering and Information Technology, Shiraz University of Technology

___

**Abstract**: The applications of Artificial Intelligence (AI) methods especially machine learning techniques have increased in recent years. Classification algorithms have been successfully applied to different problems such as requirement classification. Although these algorithms have good performance, most of them cannot explain how they make a decision. Explainable Artificial Intelligence (XAI) is a set of new techniques that explain the predictions of machine learning algorithms. In this work, the applicability of XAI for software requirement classification is studied. An explainable software requirement classifier is presented using the LIME algorithm. The explainability of the proposed method is studied by applying it to the PROMISE software requirement dataset. The results show that XAI can help the analyst or requirement specifier to better understand why a specific requirement is classified as functional or non-functional. The important keywords for such decisions are identified and analyzed in detail. The experimental study shows that the XAI can be used to help analysts and requirement specifiers to better understand the predictions of the classifiers for categorizing software requirements. Also, the effect of the XAI on feature reduction is analyzed. The results showed that the XAI model has a positive role in feature analysis.

___

## 1. Introduction

Requirement engineering (RE) is one of the main phases in the software development life cycle. RE contains different activities which are done by the requirement specifiers or software analysts. Requirement elicitation, analysis, validation, and management are usually known as the generic activities in requirement engineering [1]. The requirement elicitation activity is aimed to identify the services that should be provided by the software along with the constraints and the quality metrics that should be satisfied by the software. This activity contains different steps such as requirement discovery, requirement classification, requirement prioritization, and requirement specification.

The requirement elicitation activity has an important effect on the quality of the ultimate software. Any deficiency in this activity such as missed requirements, non-tracked requirements, requirements with contradiction, ambiguity in requirements, etc. will decrease our chance to develop high-quality software. Due to the high dependency of the quality of the software on the true requirements and because requirement elicitation is a time-consuming task, using best practices and automated solutions are suggested. As an example of automated solutions, machine learning methods such as text classifiers can be used for categorizing Functional Requirements (FR) and Non-Functional Requirements (NFR).

Usually, the software requirements are categorized as FR or NFR. FRs show the responsibilities or duties of software, and NFRs represent the quality or constraints that should be satisfied by software while performing its duties. Most of the NFRs represent the quality metrics of software such as security, performance, reliability, usability, etc.

In recent years, the impact of Artificial Intelligence (AI) on different domains has increased. Many AI solutions have been presented by researchers to solve complex problems and help people and domain experts to do better their jobs. Machine learning is a sub-domain of AI that contains different categories of supervised, unsupervised, and reinforcement methods.

In recent years, supervised algorithms have been used by researchers for requirement classification. Binkhonain and Zhao presented a systematic review of machine learning methods for the identification and classification of NFRs [2]. Their study showed that 16 different machine learning methods have been used by researchers and the supervised algorithms are the most popular. They mentioned that while the machine learning methods are successful in requirement analysis, we face some open challenges in this field that can be solved by collaboration between software engineering and machine learning researchers.

Khayashi et al. applied different deep-learning techniques to classify software requirements as functional or non-functional [3]. They applied five deep learning algorithms along with two voting methods on the PURE dataset. The results showed that applying the voting methods on the deep-learning algorithms produces better results.

Canedo and Mendes studied the combination of three feature extraction methods so-called Bag of Words (BoW), Term Frequency–Inverse Document Frequency (TF-IDF), and Chi-Squared ($CHI^2$) with four classification algorithms, Support Vector Machine (SVM), Multinomial Naive Bayes (MNB), k-Nearest Neighbors (kNN), and Logistic Regression (LR) [4]. They found that the combination of TF-IDF and LR has the best performance in binary classification.

The classification of FRs was considered by Rahimi et al. [5]. They proposed an ensemble method by a combination of five classification algorithms. In another work, they applied different deep-learning algorithms for requirement classification in one-phase and two-phase methods. First, they classified requirements as FR and NFR. Next, they sub-classify NFRs. Quba et al. studied the performance of the combination of BoW with the SVM and KNN for software requirement classification [7]. The results showed that the combination of BoW and SVM produces better results.

The previous studies showed the good performance of machine learning methods for requirement classification. Despite the successful application of machine learning, these methods are used as black-box solutions. These methods are called black boxes because their operation is not visible to the users. Assume that we have a complex supervised learning method. Due to its complexity, we can't understand how the algorithm generates a specific result. In some cases, an expert domain needs to know why the algorithm came to a result to trust the algorithm.

Here, the XAI comes into play. The XAI is a way that utilizes different methods and techniques to explain the AI models. In recent years, different XAI methods have been proposed. Local Interpretable Model-Agnostic Explanations (LIME) and SHapley Additive exPlanation (SHAP) are among the most commonly used explainers [8]. Both of them are known as surrogate models and use black-box machine learning models. They can be used for explaining different types of classifiers.

The LIME was presented by Ribeiro et al. [9]. The LIME is aimed to explain the results produced by any classification or regression model. LIME explains a prediction by approximating it locally.

Lundberg and Lee presented the SHAP algorithm [10]. Similar to LIME, the SHAP algorithm can be used to explain how each feature of a classification or regression algorithm contributes to a prediction. SHAP is an explanatory framework that uses a game theory approach and helps us to know why a machine learning model generates a specific output.

Since the introduction of XAI, this domain has received much attention from researchers. Different explanation models have been proposed and the explainers have been successfully applied to different practical problems. Although different application domains have been covered by XAI, there are a few works that considered XAI and software engineering or software requirement analysis simultaneously.

Habiba et al. presented a discussion about the synergy between XAI and requirement engineering [11]. They mentioned the challenges and proposed a framework to cope with them. The main challenges are the absence of a mediator role, no coherent definition of explainability, lack of stakeholder-centric methods, and no common vocabulary. The proposed framework to cope with the challenges contains five phases which are started with stakeholder identification, and end with requirement classification. The role of XAI in different steps of the method was discussed.

Clement et al. presented a review on the alignment of XAI and the software development process [12]. They considered the main steps in the software development process such as requirement engineering, analysis, and design, and aligned the XAI methods and tools with these development phases. They aimed to provide a roadmap for developing AI applications, utilizing explainability by considering the classic development life cycle of software systems. They suggested five research questions where each phase in software development covered one of these questions.

The supervised learning methods containing classification and regression, have been successfully applied to solving software engineering problems. However, in most cases, these methods are used as a black box and their solutions are not interpretable. Due to the importance of requirement analysis, the reliability of the black box model is very important. It seems that the XAI can help us to determine the reliability degree of the supervised methods in the classification of the requirements. Hence, this work is aimed to adapt a well-known explainer model with the software requirement classification algorithms.

This work presents a method for explaining the results generated by a requirement classification algorithm. Improving the performance of the classification method is not our focus. Hence, we use the Random Forest (RF) classifier. The method utilizes the LIME explainer along with the domain experts' knowledge to interpret the class of the requirements generated by the classifier. For this purpose, we focus on the keywords in the requirements that support or contradict the classifier's solution. This helps the expert domain to know why a requirement is classified as functional or non-functional.

The main contributions of this work are:

- The applicability of the XAI for the requirement classification step in the requirement elicitation activity is studied.
- A framework for explaining the software requirement classifier is proposed.
- A state-of-the-art explainer model is used to explain the quality of the software requirement classifier.
- The performances of the explainers are analyzed in detail.
- The results are investigated by the domain experts and their opinions are given.

The paper is organized as follows: An introduction to the XAI and LIME algorithm is given in the next section. Section 3 presents the proposed method. Section 4 reports the applicability of the model and experimental results. Finally, Section 5 concludes this work.

**2. Explainable Artificial Intelligence**

XAI is a newly emerged concept that is aimed to add explainability to AI solutions. An explainable model can be understood, trusted, and justified by human users, and provide transparency in complex AI solutions. XAI follows two objectives: creating accurate and efficient AI solutions, and providing clear explanations for their results [13].

The XAI methods can be categorized into three classes:

- Methods for interpretability: Model interpretability techniques aim to understand the internal workings of AI models and identify which features are most important in making predictions.
- Explanation generation methods: These methods are used to clearly explain the prediction of AI models.
- Sensitivity analysis methods: These methods are used to show how the model's prediction is affected by the small changes to input data.

In recent years different explanation models have been proposed by researchers where LIME and SHAP are among the most popular ones [13]. Due to the several numbers of explainer models, the selection of an appropriate explainer may be a challenging task. Jesus et al. proposed a test model that helps the user to select the appropriate model [14].

Here, we have used LIME which is suitable for our task. LIME is one of the popular explanation models and different variations of this model have been proposed in recent years. LIME is used for explaining the predictions made by machine learning models. LIME is model-agnostic that explains predictions made by any type of machine learning model. LIME perturbs the input data and analyzes the effect on the prediction. For a given prediction, LIME generates random samples around the original input. Then it fits a simple, interpretable model to these samples and their corresponding model predictions. Finally, the model is used to explain the prediction made for the original input data [9].

LIME provides local explanations. It means that it explains the prediction for a specific instance rather than a part or whole of the input. It is a useful feature, particularly for black-box models with complex internal structures. LIME can be used for classification and regression algorithms. In general, LIME is model-agnostic and provides accurate and understandable local explanations.

The LIME algorithm can be modeled as follows. For the sake of simplicity, the details are ignored here. For the full description of the LIME algorithm, the readers are referred to [9]. Assume that we have an input instance $x$ and we want to explain its prediction suggested by classifier $C$. The prediction made by the model is $y = C(X)$. LIME algorithm generates $n$ instance $x' = \{x'_1, x'_2, ..., x'_n\}$ around the original instance by perturbing it.

The output $y'_i = C(x'_i)$ is generated for the instance $x'_i$ by the black-box algorithm and the instance is labeled. These instances will be used as a dataset to train the interpretable model. Then, the LIME algorithm weights the instances according to their proximity to the original sample $x$. In this way, the instances closer to $x$ have larger weights. The weight of each sample $x'_i$ using Gaussian kernel:

$$w(x'_i) = exp\left(-\frac{d(x,x'_i)^2}{2\sigma^2}\right) \quad (1)$$

where $d(x, x'_i)$ is the Euclidean distance between $x$ and $x'$, and $\sigma$ controls the spread of the kernel. After that, the weighted average of predictions $y'_i$ is computed. Next, the weighted predictions are used to fit an interpretable model $g$. LIME employs inherently interpretable models such as Linear Regression (LR), Decision Tree (DT), and Rule-Based Heuristic (RBH) models to explain the result. Assume, LR is used. The interpretable model $g$ is fit to the perturbed sample $x'_i$ as follows:

$$g(x'_i) = \theta_0 + \sum_{i=1}^{k} \theta_i x'_i \quad (2)$$

where $\theta_i$ is the coefficient of the $i$-th feature of LR. Finally, the generated model $g$ is used to explain the prediction made by the model $C$ for the original input $x$. The explanation of the LIME is a set of feature importance scores. Each weight shows how much a feature in $x$ contributes to the prediction. For LR, the coefficients θ can be considered as the weight which are reported by the LIME algorithm.

## 3. The Proposed Method

Due to the successful applications of XAI in different domains, in this work, we try to apply the LIME explainer to the requirement classification problem. This study is aimed to answer the following research questions:

**RQ1**: Is the XAI applicable to the classification of software requirements?

**RQ2**: How can the development team members such as requirement specifiers, analysts, or architects use the XAI in the software development phase?

**RQ3**: What are the outcomes of applying XAI in the requirement engineering phase for the software development team?

To answer the research questions, a method containing the machine learning classifier and LIME explainer is adapted for analyzing software requirements. The structure of the proposed method is presented in Figure 1. The figure shows that the explanation structure has two main parts. The first part is dedicated to the classifier and the second part belongs to the explainer. To use the XAI model along with a classifier, first, the classifier is designed. Next, the XAI model can be used to explain the required instances.

The method receives a set of requirements as input. Each record in the dataset has two columns where the first column is a sentence or a paragraph of a requirement, and the second column is the class label. The requirements are given in natural language. Hence, the data should be preprocessed. The preprocessing phase contains some well-known steps that are generally used in NLP applications. The important tasks are tokenization, stop word and number removal, and stemming.

In addition to these steps, we add a new step that enables the domain expert to apply application-specific processing on the output of the stemming step. The reason for adding such a step is described in the next section. As an example, modal words such as "*shall*", "*should*", "*must*", etc. are usually used for determining the priority of requirement and it seems that they may be ignored for requirement classification. Hence, these words can be removed from the requirements. Also, the name of the software may not provide the extra information for requirement classification.

After preprocessing we are ready to move to the next phase. The output of the preprocessing phase is a set of vectors where each vector contains the roots of important words in a requirement along with its class label. These vectors are used as the input for the next phase of the classification part.

Most classifiers don't work with textual data as the input. Therefore, the textual data is usually converted to the vectors of numbers using different approaches. Here, we used the TF-IDF method to convert the requirements to numerical vectors. After applying the TF-IDF method, the classifier is used to classify the requirements. By training the classifier, the generated model can be used by the XAI model to explain why an algorithm is classified as functional or non-functional.

Here we use the LIME explanation model. The LIME receives the requirement we want to explain and the generated classifier as its input and produces the explanation. For textual data such as user requirements, the LIME model generates a set of supportive and contradictor words along with their weight in the explanation and the probability of predicting the class of the requirement.

Finally, the output of the XAI model along with the class predicted by the classifier can be used by domain experts such as analysts, requirement specifiers, or architects to use them as a part of their works. Then the domain expert analyzes the predicted class labels along with the explanations generated by the LIME algorithm. The outcome of the analysis can be used as feedback for the preprocessing phase. For example, the expert may remove a feature (he/she thinks that) doesn't important for requirement classification and regenerate new results and see the effect of its knowledge on the performance of the algorithm. The feedback can acts as a feature reduction mechanism.

It seems that the proposed method helps the domain experts in the software development process to see the effective features in the prediction of a requirement's class. Hence, domain experts can trust the output of the machine learning models in requirement analysis.

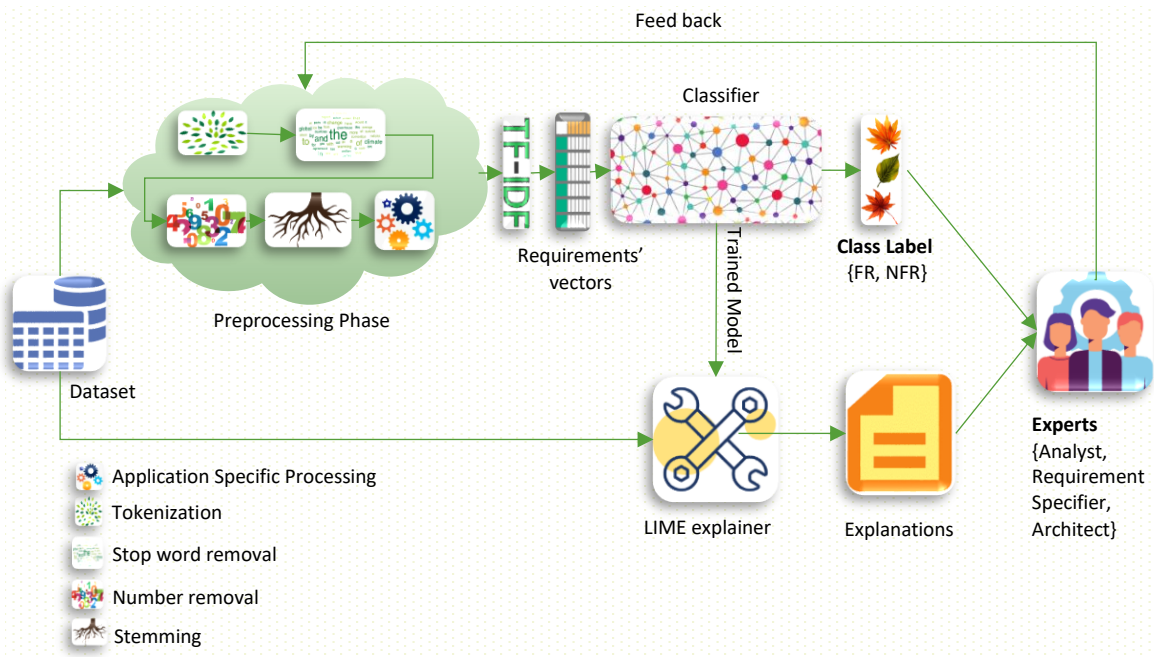

Figure 1. The structure of the proposed method

## 4. Experimental Study

To study the applicability of the proposed method containing the XAI module for software requirement analysis, the PROMISE dataset is used. The dataset contains functional and non-functional requirements. The FRs have the label (F), while the NFRs are divided into eleven categories with the following labels: Availability (A), Fault Tolerance (FT), Legal (L), Look & Feel (LF), Maintainability (MN), Operational (O), Performance (PE), Portability (PO), Scalability (SC), Security (SE), and Usability (US). In this work, we considered requirement classification as a binary classification problem. Hence, all the sub-categories of NFRs are considered non-functional. Roughly, 40% of the data is functional, and 60% is non-functional.

We didn't do any upsampling or downsampling on the data. The PROMISE dataset is partitioned into a train set and a test set where 80% is used for training and the remaining ones are used for testing. The samples in the train and test sets are selected randomly. The method was implemented using Python programming languages. The NumPy, NLTK, Keras, Lime, Pandas, and Seaborn packages were used.

The LIME algorithm can work with any classification method. So, any machine learning or deep learning algorithm can be put as the classifier module in

Figure 1. Here, we used the RF as the classification algorithm. The performance of the classifier on the test data is given in Table 1. The average results after 30 runs of the method are reported in this table. The results show that the Random Forest has acceptable performance on the test data.

Table 1. Performance of the Random Forest on the test set

| Metric | Value |
|---|---|
| F1-score | 0.87 |
| Accuracy | 84.08% |
| Precision | 0.84 |
| Recall | 0.91 |

### 4.1. Explainer Output

Before we analyze the applicability of the LIME algorithm for requirement analysis, the output of the LIME is presented here. The solution that is generated by LIME contains the set of words that support or distract the decision. An example of the LIME output is given in Figure 2. The vertical axis shows the list of supportive and distractive words. The supportive and distractive words are shown in green and red colors respectively. The horizontal axis shows the weight of support or distraction of the words. Also, the LIME provides the probability of considering a requirement as functional or non-functional.

The corresponding requirement is given in Figure 3. The text of the requirement after preprocessing is shown here. This is an NFR and the probability of non-functionality by the LIME is larger than 0.9. This figure shows that some of the words such as "*application*" and "*after*" have not been considered to classify this requirement.

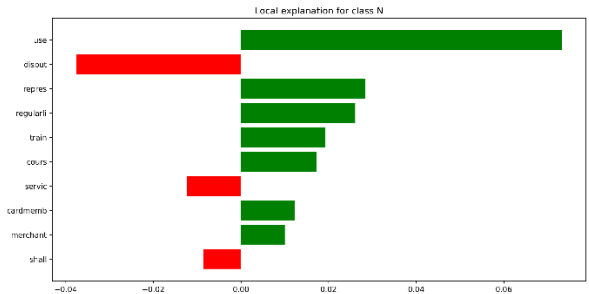

Figure 2. The output of the LIME explainer on a randomly selected requirement in the PROMISE dataset.

100% of the cardmember and merchant services representatives shall use the Disputes application regularly after a 2-day training course.

Figure 3. The corresponding requirement for the explanation of Figure 2.

### 4.2. The Model's Explainability

This experiment is aimed to analyze the explanation of the NFR predictions in the dataset using the LIME algorithm. It seems that the explainer output as shown in Figure 2 provides valuable insight for us to analyze the applicability of the LIME algorithm. The results that are reported here and in the next section only consider the test sets.

To analyze the applicability of the LIME for requirement analysis, two sets of words that are reported by the LIME algorithm are used for determining requirements as non-functional. Here we only consider the explanation of a requirement as NFR. The same scenario can be used for functional requirements. These words are known as supportive (S) and distractive (D) words. The definitions of the supportive and distractive words are given in Table 2.

Table 2. Definition of the supportive and distractive words

| Metric | Definition |
|---|---|
| Supportive words (S) | Set of all words with a positive effect in classification a requirement as NFR. |
| Distractive words (D) | Set of all words with a negative effect in the classification of a requirement as NFR. |

Three experiments are done to show the explainability potential of the LIME algorithm for the software requirements. In the first experiment, all the supportive and distractive words in the requirements are considered. For this purpose, the test data is given to the LIME algorithm sample by sample, and the supportive and distractive words are collected. All of these words are aggregated and their occurrences are computed.

Figures 4 and 5 show all the supportive and distractive words to explain the classification of a requirement as non-functional. The horizontal axis shows the words that appeared in NFRs, and the vertical axis shows the percentage of occurrences of the words. The percentage of occurrences is defined as follows:

$$pc_i = \frac{n_i}{N} \times 100 \qquad (3)$$

where $n_i$ is the number of occurrences of the word $i$, and $N$ is the total number of words. Dozens of supportive and distractive words are reported by the LIME algorithm for explaining all the test samples. It is not possible to show all of them. Hence, only the top 30 words with the higher percentage of occurrences are shown here. The roots of the words are presented in these Figures.

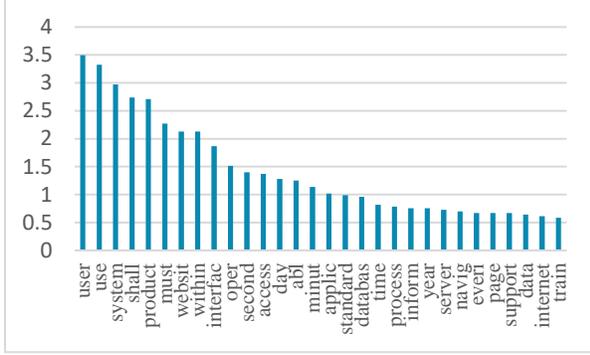

Figure 4. List of most repetitive and supportive words in classifying a requirement as non-functional in the dataset before removing modal words.

As shown in Figure 4, "*system*", "*user*", "*product*", "*use*", and "*shall*" are the five supportive words with the highest repetitions. Although these words are highly repeated in predicting a requirement as an NFR, usually a human expert cannot determine an NFR by using only these words, and to classify a requirement as NFR, he/she needs more precise words. Most likely we have the same situation in the machine learning model. It seems that the general words in combination with the specific NFR's words are used by the machine learning methods for classification.

Also, "*shall*", "*product*", "*user*", "*system*", and "*dispute*" are the five distractive with the highest repetition as shown in Figure 5. Some of the words with high repetition such as "*shall*" and "*product*" are seen both in supportive words and distractive words.

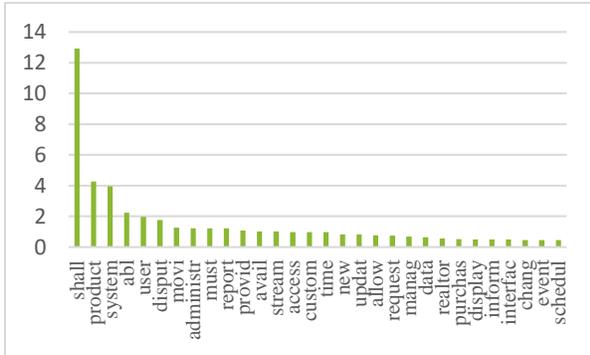

Figure 5. All the distractive words in classifying a requirement as non-functional in the dataset before removing modal words.

To better discriminate, and analyze the words, we partitioned all the words used by the explainer to explain the RF classifier into three sets. The sets are $A$, $B$, and $C$, and are defined as follows:

- $A = D - C$: is a set of distractive words minus the common words.
- $B = S - C$: is a set of supportive words minus the common words.
- $C = S \cap D$: is a set of common words which is the intersection of sets S and D.

Figure 6 shows these three sets generated by the LIME explainer. Only, the top 30 words with the highest repetition are shown in this figure. Here, we have the following members in each set:

$A = \{allow, dispute, diplay, request, ...\}$ (4)
$B = \{operate, second, minute, navigation, ...\}$ (5)
$C = \{product, shall, system, user, ...\}$ (6)

The categorization of the words shows that the explainer reasonably uses them to predict a requirement as functional or non-functional. The sets $A$ and $B$ are more important than set $C$. However, the common words need to be considered in more detail.

The set $C$ of common words contains words like "*shall*", "*must*", "*user*", etc. Although these words are highly repetitive, it seems that such words can't be significantly used by a classifier for classifying a requirement as functional or non-functional. The same situation exists in the real world, and a human expert cannot only rely on common words to classify a software requirement.

The set $A$ of only distractive words contains words like "*display*", "*change*", "*class*", "*delete*", etc. Usually, such words are used to show the responsibility or services provided by the software. These words are candidates to determine a requirement as FR. Hence, it seems that the explainer reasonability identified them as distractive words. It should be noted that due to classification errors, the categorization of the words in these sets may contain errors.

The set $B$ of only supportive words contains words like "*second*", "*minute*", "*install*", "*license*", "*threat*", etc. which are reasonable for classifying requirements as non-functional. They are interesting words that are used by the classifier to classify a requirement as non-functional. For example "*second*" is a time unit and the users use this word to express maintenance or performance metrics such as "*Response Time*" or "*Fix Response Time*". The word "*install*", shows that it is used for the operational or maintainability of the software, and we expect to face such a metric.

As an example, two NFRs containing the words "*second*" and "*install*" in the dataset are shown in Figure 7. Usually, in the real world, a human expert uses such keywords to categorize a requirement as NFR. It seems that the model takes such words as a part of its decision-making process to predict a requirement as NFR. In the dataset, most of the occurrences of this word belong to the NFRs.

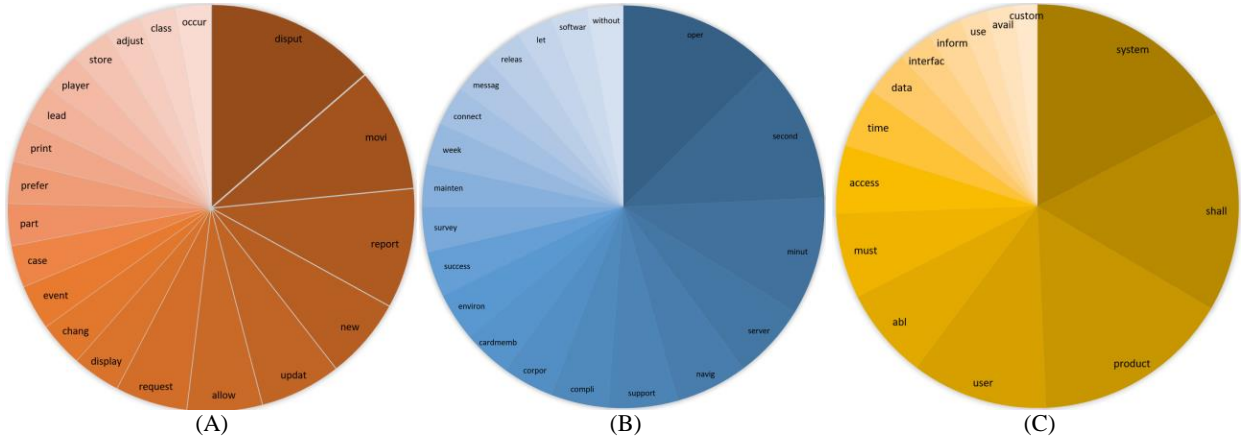

Figure 6. Three sets of words (A) distractive words, (B) supportive words, (C) common words

**Req 1**: When purchasing a streaming movie or pre-paid card via credit card the processing time should have a maximum response time of 15 seconds.

**Req 2**: The product will function alongside server software on any operating system where the Java runtime can be installed.

Figure 7. Sample requirements are correctly classified as NFR by the classifier.

The list of supportive words generated by the LIME algorithm is analyzed by human experts. The experts aimed to extract a set of words that are usually used in the real world to categorize a requirement as NFR. Table 4 shows a portion of these words that are used by the proposed model for predicting a requirement as NFR.

Table 4. A subset of words used by the proposed method for predicting requirements as NFRs

| Word | NFR class |
| --- | --- |
| Available, Failure | Availability |
| Operate | Fault Tolerance |
| Laws | Legal |
| Navigation, Resolution | Look & Feel |
| Budget, | Maintainability |
| Install, License | Operational |
| Second, Maximum, Load | Performance |
| Operating | Portability |
| Support, Concurrent, Increase | Scalability |
| Logon, Threat | Security |
| Understand, Screen, Comfort | Usability |

Some words are shared among more than one sub-class of NFRs. However, each word is only shown in one sub-class of NFR where it has the highest probability of occurrences. This experiment shows that the classifier almost works similarly to a human expert in determining a requirement as NFR. In comparison to general words such as "*product*" or "*user*", these specific words have lower frequencies. So, in the figures, most of these words don't appear. But, they are successfully used by the model to predict a requirement as NFR.

**4.2. XAI for Feature Analysis**

The analysis of the words recommends that the common words may be considered in more detail to see their effect on classification. Here, we examine a part of the common words which are known as modal words, and the top 3 common words (e.g. "*system*", "*product*", "*user*"). In requirement elicitation, modal words such as "*shall*", "*should*", "*must*", etc. are usually said by the users to describe the degree of importance of a requirement. Such keywords are useful for requirement prioritization. However, in requirement classification, it seems that these modal words cannot be used independently for classifying a requirement.

These words appear as supportive and distractive because they have a high frequency of occurrence in the requirements. Hence, we study the effect of removing these words as a part of the preprocessing phase for software classification applications. The feedback line in Figure 1 may be triggered in such cases. The feedback is suggested here to apply the insights gathered from the output of the model on the preprocessing phase and provide the ability for the domain expert to see the effects of his/her changes on the input data.

Also, the other three words "*system*", "*user*", and "*product*" are general words in the requirement engineering phase, and they don't specifically determine the class of a requirement. As suggested by the LIME algorithm as the top common features, we remove them and evaluate the performance of the RF classifier for feature reduction.

In this experiment to show the applicability of XAI for feature analysis, the modal words and the top three common words are removed from the requirements. We used the t-test to compare the performance of the classifier before and after removing these words. The results of the classifier after 30 runs are used for the test. the t-test is a statistical test used to see if the performance of the RF classifier is significantly changed after the word removal or not. The significance level is set at 0.05.

The results are shown in Table 5. The A, A-M, and A-M-C represent all words, words after removing modal words, and words after removing modal and top three common words. If the results of the RF algorithm before removal are significantly better than its performance after removal (or vice-versa), the corresponding cell is set at S, else it is set as N. It is apparent that the performance of the RF classifier has not significantly changed after feature reduction. It seems that the role of the LIME algorithm for feature reduction (in this case removing unimportant features) is positive.

Table .5 t-test results on the performance of the RF classifier on the test set before and after removing modal words, and after removing top common words.

| Metric | Data | A | A-M | A-M-C |
|---|---|---|---|---|
| F1-score | A | N | N | N |
|  | A-M | N | N | N |
|  | A-M-C | N | N | N |
| Accuracy | A | N | N | N |
|  | A-M | N | N | N |
|  | A-M-C | N | N | N |
| Precision | A | N | N | N |
|  | A-M | N | N | N |
|  | A-M-C | N | N | N |
| Recall | A | N | N | N |
|  | A-M | N | N | N |
|  | A-M-C | N | N | N |

Figures 8 and 9 show the supportive, and distractive words in determining requirements as non-functional in the dataset after removing the modal and common words. These graphs show the names of keywords and the percentage of their repetition in the requirements. The number of supportive and distractive words is much more than the words shown in these Figures. Only, words with more probability of occurrences are shown here. Here, the roots of keywords are presented.

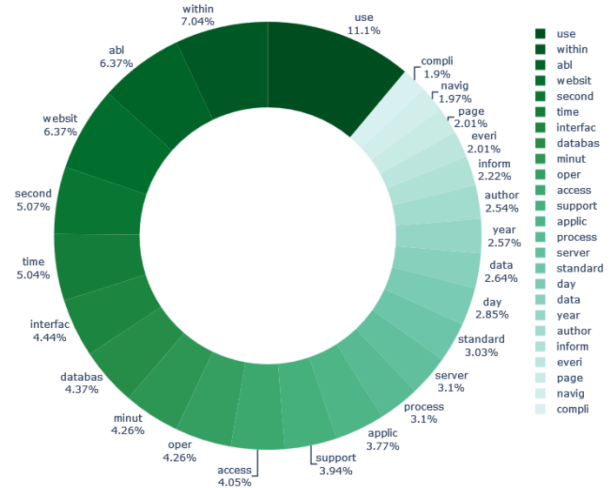

Figure 8. List of supportive words suggested by the LIME algorithm in classifying a requirement as non-functional in the dataset after removing the modal and common words.

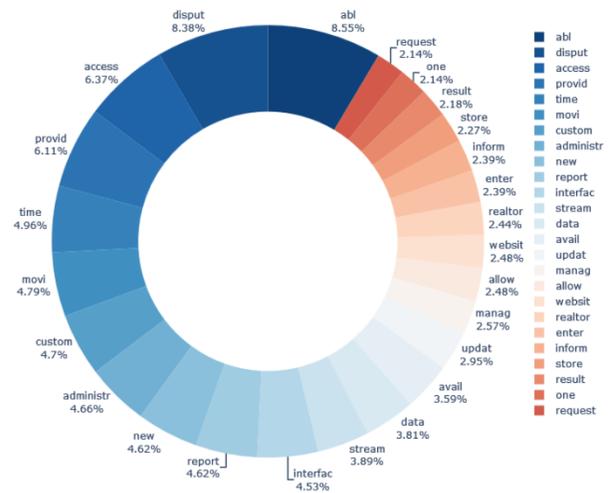

Figure 9. List of all distractive words suggested by the LIME algorithm in classifying a requirement as non-functional in the dataset after removing modal and common words.

### 4.5. Applicability Analysis

By considering the results obtained by applying the LIME algorithm to the dataset, it can be seen that the classifier uses almost relevant features (e.g. words) for requirement classification. However, some common words such as "*shall*", "*user*", "*system*", and "*products*" act as both supportive and contractive for classifying a requirement as NFR. It seems that these types of words in combination with the other supportive words classify a requirement as non-functional. Also, the results show some verbs which usually used to specify the functionality (or duty) of a system. These verbs belong to the distractive set and it seems to be reasonable.

In general, the experiments show that the XAI is applicable for requirement analysis and especially the requirement classification problem (RQ1). However, in this work, only basic analysis is considered, and it is possible to do a more detailed and complex analysis. The team members who have a responsibility in RE such as requirement specifiers, analysts, and architects can use the XAI explainers alongside the classifiers as shown in Figure 1, or in other similar ways (RQ2).

The XAI provides insight for the team members about the predictions that are generated by machine learning models. The outputs of the XAI help them to know how the machine learning models generate solutions. In the case of requirement classification, the XAI helps the team members to see the supportive, distractive, and common words in classifying a requirement as FR, NFR, or a sub-class of NFR (RQ3).

The preprocessing phase has an important role in the performance of a machine learning model. Recently, using XAI as a part of preprocessing phase has been recommended [15]. Using XAI as a way for feature selection in requirement engineering is recommended. The XAI can be used to analyze the impact of each feature and help the expert to select the feature in a better way.

## 5. Conclusion

In this work, we studied the applicability of the XAI in explaining the behavior of the requirement classification algorithms. The LIME algorithm analyzed the results produced by the classifier on the PROMISE dataset. This study showed that the XAI has a positive role in better understanding the behavior of classifiers when they are applied to textual requirements. The most important words, which are used by the classifiers in categorizing the requirements, were identified. The investigation and analysis of these words by the expert domains (e.g. analysts, architects, and requirement specifiers) showed that the suggested words are meaningful.

In this work, the explanation of machine learning methods for requirement classification was considered. Recently, different deep-learning methods have been used for requirement classification. The explanation of these methods in requirement engineering domains may show interesting results because of the good performances of such methods in classifying requirements.

Different types of explainers have been designed by researchers that can be used in this domain. In this work, only the applicability of LIME as an explainer was considered. A comparison study among these explainers is recommended. Also, the XAI as a tool for feature selection is recommended.